\documentclass[conference]{IEEE/IEEEtran}
\usepackage[utf8]{inputenc}           %
\usepackage{amsmath,amssymb,amsfonts}
\usepackage{graphicx}
\usepackage{textcomp}
\usepackage[pdftex,svgnames]{xcolor}
\usepackage{balance} %
\usepackage{tabularx}
\usepackage{paralist}
\usepackage{booktabs}
\usepackage{comment}
\usepackage[T1]{fontenc}
\usepackage[T1,lowtilde]{url}
\usepackage[abbreviate=true, dateabbrev=true, isbn=true, doi=true, urldate=comp, url=false, maxbibnames=9, backref=false, backend=biber, style=numeric-comp, language=american, giveninits=true, sortcites=true, sorting=none]{biblatex}

\usepackage{citation-cleaner}
\AtEveryBibitem{\clearlist{location}} %
\AtEveryBibitem{\clearfield{series}} %
\AtEveryBibitem{\clearlist{language}} %
\AtEveryBibitem{\clearfield{note}} %

\usepackage[hidelinks]{hyperref}
\usepackage[capitalize,noabbrev]{cleveref}
\usepackage[shortlabels,inline]{enumitem}
\usepackage{float}
\usepackage{array}

\newcommand{\head}[1]{\par\noindent\textbf{#1:}\space}

\title{CHESS: A Framework for Evaluation of Self-adaptive Systems based on Chaos Engineering}

\author{%
	\IEEEauthorblockN{%
		Sehrish Malik\IEEEauthorrefmark{1}\quad\quad%
		Moeen Ali Naqvi\IEEEauthorrefmark{1}\quad\quad%
		Leon Moonen\IEEEauthorrefmark{1}\IEEEauthorrefmark{4}\\[0.8ex]}
	\IEEEauthorblockA{\IEEEauthorrefmark{1}Simula Research Laboratory, Oslo, Norway \quad\quad \IEEEauthorrefmark{4}BI Norwegian Business School, Oslo, Norway}
	\IEEEauthorblockA{Email:
		\{sehrish, moeen\}@simula.no, 
		leon.moonen@computer.org}%
}

\makeatletter
\def\ps@IEEEtitlepagestyle{%
  \def\@oddfoot{\mycopyrightnotice}%
  \def\@evenfoot{}%
}
\def\mycopyrightnotice{%
  \hspace*{3mm}\includegraphics[width=2cm]{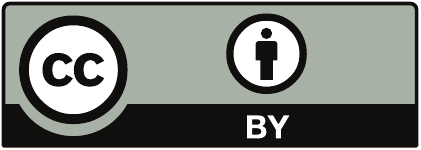}%
  \hspace*{2mm}\raisebox{2.5mm}{%
          \parbox{\columnwidth}{\footnotesize This work is licensed under a Creative Commons \\ Attribution 4.0 International (CC BY 4.0) license.}%
          \hspace*{-68pt}\mbox{1}\hspace{12pt}\fbox{\parbox{.94\columnwidth}{\footnotesize\textsl{Accepted for publication in the 18nd Symposium on Software Engineering for Adaptive and Self-Managing Systems (SEAMS 2023).}}}%
  }%
  \gdef\mycopyrightnotice{}%
}
\makeatother

\begin{document}

\maketitle
\pagestyle{plain}

\begin{abstract}

  There is an increasing need to assess the correct behavior of self-adaptive and self-healing systems due to their adoption in critical and highly dynamic environments.
  However, there is a lack of systematic evaluation methods for self-adaptive and self-healing systems. 
  We proposed \emph{CHESS}, a novel approach to address this gap by evaluating self-adaptive and self-healing systems through fault injection based on chaos engineering (CE). 

  The artifact presented in this paper provides an extensive overview of the use of \emph{CHESS} through two microservice-based case studies: a smart office case study and an existing demo application called \emph{Yelb}. 
  It comes with a managing system service, a self-monitoring service, as well as five fault injection scenarios covering infrastructure faults and functional faults. 
  Each of these components can be easily extended or replaced to adopt the \emph{CHESS} approach to a new case study, help explore its promises and limitations, and identify directions for future research.

\end{abstract}

\section{Introduction}
\label{sec:introduction}

\noindent 
Self-adaptive systems (SAS) and self-healing systems (SHS) are becoming increasingly important in fields such as the Internet of Things, Industry 4.0, and smart cities~\cite{wong2022:selfadaptive}.
These systems are designed to operate in highly dynamic environments and are expected to handle uncertainty and unanticipated behavior, as well as provide fault tolerance and resilience. 
However, due to the complexity and dynamic nature of these systems, it is challenging to anticipate all possible scenarios that these systems will encounter.
This is particularly important for the evaluation of these systems, which requires assessing the correct behavior of these systems. 

There has been a growing concern among researchers about the lack of systematic evaluation for SAS and SHS~\cite{ gerostathopoulos2021:how, passini2022:design, desousa2019:quality}.
A recent systematic mapping study found that only a small percentage of studies in this field 
focus on evaluating previously developed applications~\cite{passini2022:design}. 
In addition, there are limited tools available to support evaluations based on runtime measures, 
with most studies focusing on evaluating the models used to design the system~\cite{desousa2019:quality}. 
These models, however, may not take into account all potential scenarios 
that a system may encounter during operation. 
While runtime models offer a solution to these limitations, 
they also come with their own challenges, such as maintenance and the need for model creation~\cite{ghahremani2020:evaluation}.
Hence, there is a need for mechanisms to evaluate SAS and SHS based on runtime measures 
that consider potential scenarios a system may encounter during operation. 

To fill this gap, our earlier work proposed \emph{CHESS}, 
an approach for the systematic evaluation of self-adaptive 
and self-healing systems that build on chaos engineering principles~\cite{naqvi2022:evaluating}. 
\emph{CHESS} systematically perturbates the system-under-evaluation 
and records how the system responds to those perturbations.
The artifact\footnote{~CHESS artifact on Zenodo: \url{https://doi.org/10.5281/zenodo.6817763}\label{artifact}} presented in this 
paper 
provides an extensive overview of the use of 
\emph{CHESS} through two microservice-based case studies: 
a smart office case study and an existing demo application called \emph{Yelb}. 
Concretely, the artifact consists of 
(i) predefined functional and infrastructural level fault injection scenarios,
(ii) a self-monitoring service that presents extensive logs 
for the deployed services' normal and abnormal behaviors,
(iii) the managing system service that reacts to the system's abnormal behavior traces and brings the system back to a stable condition, 
and (iv) a comparison of the service failure and cascading effects with and without deployment 
of the managing system service. 

The remainder of this paper is organized as follows.
In \cref{sec:background}, we summarize self-adaptive and self-healing systems for the microservice architecture, evaluation of these systems based on CE, and position \emph{CHESS} in the landscape of self-adaptive system artifacts. 
\cref{sec:approach} presents the design and architecture of \emph{CHESS}, along with the microservice-based case studies and test scenarios considered for the fault injection.
\cref{sec:use_artifact} describes how to use the artifact with both of the case studies and presents the results of the fault injection scenarios. 
We conclude in \cref{sec:conclusion}, including discussions of the artifact's applicability and directions for future work.

\section{Background and Positioning of the Artifact}
\label{sec:background}

\noindent
We briefly introduce the basics of SAS and SHS, and their evaluation in the context of microservices architecture, and highlight how \emph{CHESS} complements 
the existing artifacts for engineering self-adaptive systems. 

\subsection{Self-Adaptive System for Microservices Architecture}

\noindent 
Self-adaptive systems are a class of software systems that have the ability to automatically adapt to changes in their environment~\cite{weyns2021:introduction}.
At a high level, these systems can be seen as comprising a \emph{managed system} that is controlled by a \emph{managing system}, generally realized through a MAPE-K feedback loop~\cite{kephart2003:vision}. 

Self-adaptive systems for microservices architecture can bring multiple benefits, 
such as increased scalability, improved reliability, and reduced maintenance costs.
Studies have suggested that there is an overall improvement in the management of microservices-based applications,
allocation of microservices among the available servers,
quality attributes such as performance, scalability, and resilience 
through the introduction of self-healing, self-management, 
and self-optimization properties~\cite{filho2021:selfadaptive}.
Furthermore, an architecture-based self-adaptation framework with a MAPE-K feedback loop 
for a microservice as a \emph{managed system} shows a reduced cost of ownership 
and faster self-adaptation~\cite{boyapati2022:selfadaptation}. 
On the other hand, the introduction of microservices architecture, as a \emph{managing system}, can improve the self-adaptation 
capabilities of systems for various measures such as run-time data analysis~\cite{banijamali2020:kuksa}.
Some challenges in developing a self-adaptive system based on microservices include
developing monitoring and adaptation mechanisms for ensuring quality attributes;
determining the level of distribution, observability, 
and granularity for deploying \emph{control components}; 
and determining mechanisms for evaluation of the given quality attributes~\cite{mendonca2021:developing}.

\subsection{Evaluation of SAS and SHS based on Chaos Engineering}

\noindent 
The evaluation of self-adaptive and self-healing systems is an important aspect of their design and deployment.
In our previous work~\cite{naqvi2022:evaluating}, we defined the evaluation of self-adaptive and self-healing systems as
"\emph{an approach to determine if a system meets objectives under operation, identify areas in which the system performs as well as desired or predicted, and provide evidence to the value and applicability of the system}." 

Several approaches to the evaluation of SAS and SHS include model-based evaluation~\cite{barbosa2017:lotus}, 
metric-based evaluation~\cite{porter2018:tess}, model checking~\cite{camara2012:evaluation}, 
and runtime testing and verification~\cite{king2011:comparative} each with its benefits and limitations. 
However, none of these evaluation approaches are viable for evaluation 
covering the execution of the system under real-life failure scenarios.
Therefore, to address this gap, we introduce a mechanism that builds on \emph{chaos engineering} principles.
Chaos engineering (CE) is the practice of intentionally causing and studying controlled chaos 
within software systems operating in realistic environments, 
with the goal of increasing the systems' resilience and ability to handle unforeseen circumstances~\cite{basiri2016:platform}.
The core tenets of CE can be outlined as four main principles, 
which include formulating hypotheses based on the steady-state behavior of systems, 
introducing variations to real-world events, 
conducting experiments in a production environment, and automating these experiments for continuous execution.
Our approach evaluates the systems through a systematic process that involves exposing the system to faults and testing its ability to recover from such perturbations.

\subsection{Positioning of the Artifact}

\noindent 
Artifacts play a vital role in advancing the field of self-adaptation. 
They serve as tangible examples of the algorithms and techniques developed by researchers, 
allowing for their evaluation and assessment. 
In addition, artifacts provide problematic scenarios and solutions that can inspire further research, 
as well as facilitate the comparison of results among different studies.
The self-adaptive exemplars website\footnote{~\url{http://self-adaptive.org/exemplars/}} gives an overview of re-usable artifacts produced by researchers and engineers in the self-healing and self-adaptive community.

Various existing artifacts focus on web-based systems, microservices architecture, or cloud environments that assist in the evaluation of the \emph{managed systems}. 
Znn.com~\cite{cheng2009:evaluating} is a web-based information system that mimics 
real-world systems and provides an experimental environment to facilitate the evaluation. 
The exemplar applies a self-adaptive framework, \emph{Rainbow} and presented an 
evaluation of the self-adaptive system based on a benchmark.
Hogna~\cite{barna2015:hogna} is a platform for deploying self-managing 
web applications on the cloud. 
It automates operations, monitors the health of the applications, extracts metrics, 
and analyzes performance data based on a model to create and execute an action plan. 
K8-Scalar~\cite{delnat2018:k8scalar} is an exemplar that allows the evaluation of different self-adaptive approaches to autoscaling container-orchestrated services.
It is based on Docker, and Kubernetes, and extends a generic testbed for scalability evaluation of large-scale systems called Scalar.
SEAByTE~\cite{quin2022:seabyte} enhances the automation of continuous A/B testing of a 
microservice-based system. 
Furthermore, exemplars such as SWIM~\cite{moreno2018:swim}, DARTSim~\cite{moreno2019:dartsim}, 
and RDMSim~\cite{samin2021:rdmsim} consist of \emph{managed systems} 
which can assist in the evaluation of external adaptation managers.

The present artifact represents a departure from existing artifacts in the field, 
as it prioritizes the \emph{evaluation} of \emph{managing systems} 
by inducing faults within the \emph{managed system}. 
Thus, as opposed to primarily focusing on the \emph{creation} of new self-adaptive approaches, 
it supports one of the critical tasks of software engineering research, i.e., the \emph{systematic evaluation} of novel approaches.
This aligns with the growing desire to produce artifacts in self-adaptation
that support industry-relevant research~\cite{weyns2022:guidelines}.

\begin{figure}
	\vspace*{1ex}
	\centering
	\includegraphics[width=0.90\columnwidth]{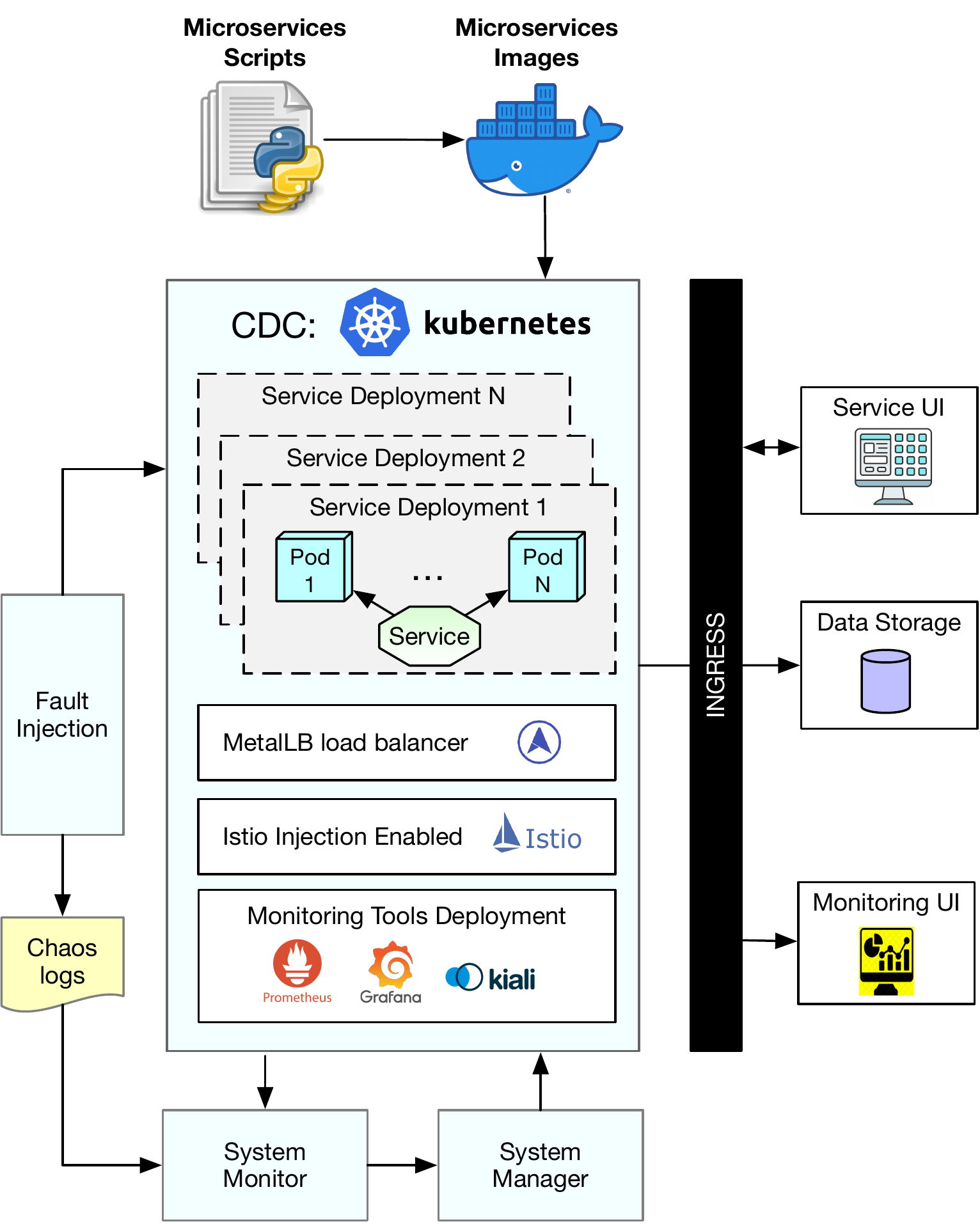}
	\caption{\label{fig:dep1}Implementation Architecture for the applications deployment}
	\vspace*{-2ex}
\end{figure}

\section{CHESS}
\label{sec:approach}

\noindent
In this section, we present the architecture of CHESS, the demo applications used for testing, and the test scenarios for fault injection. 

\subsection{Design and Architecture}

\noindent
Figure~\ref{fig:dep1} presents a detailed architecture for the CHESS approach. It consists of four main modules: containerized deployment cluster (CDC), system monitor, system manager, and fault injection. These modules are presented below.

\head{Containerized Deployment Cluster (CDC)}
This artifact discusses the process of introducing faults into microservice-based applications deployed in a containerized environment using Kubernetes (K8s). 
The environment details are shown in Table~\ref{tab:table-k8s}.
The K8s cluster starts with a total of 20 GB memory and 4 CPUs to ensure that it has sufficient resources to run the demo applications.
To make the demo applications accessible from outside the cluster, we have configured the K8s cluster with MetalLB, which is an addon that enables external IP services in K8s. 
In addition, the Istio service mesh is installed in the cluster to monitor the microservices traffic flow and to provide an easy way to manage the traffic between microservices.
In order to visualize the data traffic and services' status, we have installed Grafana, Kiali, and Prometheus monitoring tools. 
Grafana is used to visualize the data traffic, while Kiali is used to visualize the services' status. 
Prometheus is used to query data metrics from its database and to store the data for later analysis. 
The combination of these tools provides a complete solution for monitoring microservices-based applications and for evaluating the fault tolerance of these applications.

\begin{table}[b]
	\vspace*{-2ex}
	\caption{\label{tab:table-k8s} K8s Cluster Details}
	\centering
	\begin{tabular}{ll}
		\midrule
		K8s cluster &     Minikube    \\
		Cluster memory &      20 GB   \\
		Cluster cpus  &  4 \\
		Load balancer & MetalLB \\	
		Service mesh & Istio  \\
		Monitoring DB & Prometheus \\
        Monitoring tools & Grafana \& Kiali \\
		Chaos tools & chaostoolkit-Kubernetes \\
 		\bottomrule
	\end{tabular} 
\end{table}

\head{System Monitor}
The system monitor module monitors the status of the running services and performs necessary checks to ensure their validity. 
Its purpose is to observe the behavior of the system under various conditions, distinguish the system's normal behavior patterns from abnormal ones, and alert the system manager in case of any abnormal behavior.

\head{System Manager}
The system manager module plays a crucial role in ensuring the continuous operation of services deployed in the CDC. 
The primary responsibility of the system manager module is to receive abnormal system behavior alerts from the system monitor module and handle the recovery phase of the services that encounter faults.
In this artifact, the system manager adopts a rule-based approach for fault recovery. 
The recovery process is automated and follows a set of predefined rules, which have been configured for the selected demo applications.
The use of the system manager module also enables us to compare the impact of faults, known as the blast radius, for different fault injection scenarios, both with and without the system manager. 
This helps us to evaluate the effectiveness of the system manager in mitigating the effects of faults in the running services.

\head{Fault Injection}
The fault injection module follows the principles of Chaos Engineering (CE) to induce faults in the deployed application services. 
A set of chaos experiments is carefully designed and scripted, for each fault injection scenario, in order to inject the desired faults into the services. 
A chaos experiment template is shown in Table~\ref{tab:table-chaos}. A chaos experiment first defines the steady state with a probe function and a check against a tolerance. 
Once the steady state is met, the chaos action method is called to inject the respective fault, using the \textit{service\_name} and \textit{namespace} arguments to identify the service.
A pause before or after the fault injection can also be added.
A virtual environment in Python is prepared with "chaostoolkit" and "chaostoolkit-Kubernetes" libraries to execute the chaos injection scripts. Each chaos injection generates chaos logs, which can be used for system observation purposes.

\begin{table}
	\caption{\label{tab:table-chaos} Chaos Experiment Template}
	\centering
	\begin{tabularx}{0.95\columnwidth}{ll}
		\midrule
		steady-state-hypothesis: &         \\
		type & probe \\
		provider type & python \\
		provider module & chaosk8s.probes \\
		func & \{deployment\_available\_and\_healthy\},  \\
             &  \{battery\_charged\}, \{timely\_response\} \\
		arguments & service\_name \\
							& namespace \\
		tolerance  &  true \\
		\hline
		chaos-method: &  \\ 
		type & action \\
		func & \{inject\_fault\}, \{deprecate\_battery\},  \\
             & \{inject\_delay\}, \{terminate\_pods\}, \\
             & \{load\_service\} \\
		arguments & service\_name \\
                            & namespace \\
        \hline
        pause & \{before, after\} in seconds \\
		\bottomrule
	\end{tabularx} 
    \vspace*{-2ex}
\end{table}
  
\subsection{Demo Applications for Testing}
 
\noindent
The artifact uses the smart office case study from our earlier paper~\cite{naqvi2022:evaluating} and an open source example application named Yelb\footnote{~The Yelb application was reused from \url{https://github.com/mreferre/yelb}}
as the demo applications for chaos injection and testing.

\begin{figure}
	\vspace*{1ex}
	\centering
	\includegraphics[width=0.90\columnwidth]{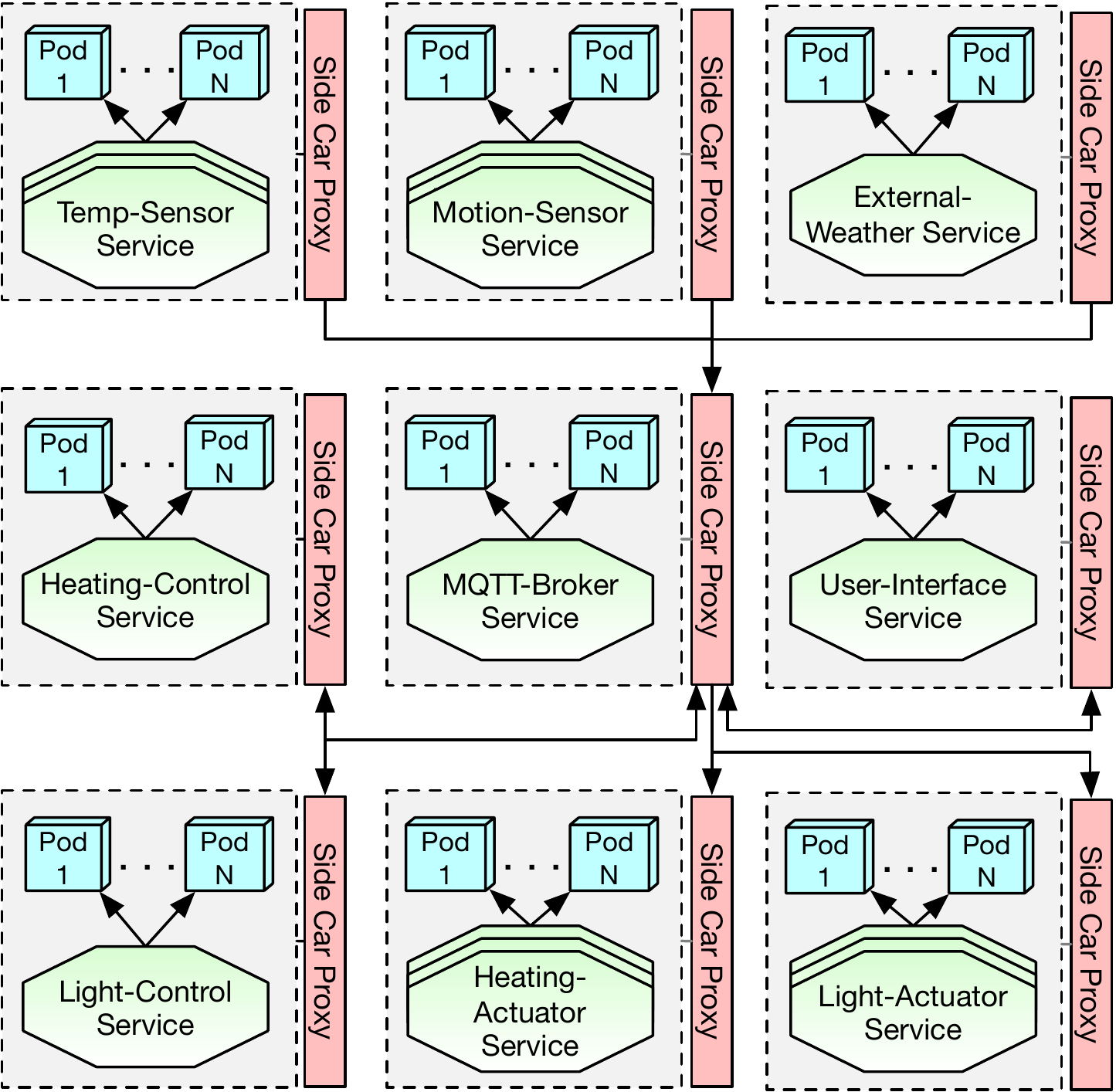}
	\caption{\label{fig:office}Service graph for the Smart Office case study}
	\vspace*{-1ex}
\end{figure}

\head{Smart Office Case Study}
The smart office case study consists of nine services, including three input services (temperature sensor, motion sensor, and external weather), two control services (heating control, light control), two actuator services (heating actuator, light actuator), an MQTT broker, and a user interface. 
The control services retrieve periodic sensing data and weather data, then use rules to control heating and lighting actuators. 
The service graph for the smart office case study is shown in Figure~\ref{fig:office}.

\begin{figure}[b]
	\vspace*{-2ex}
	\centering
	\includegraphics[width=0.99\columnwidth]{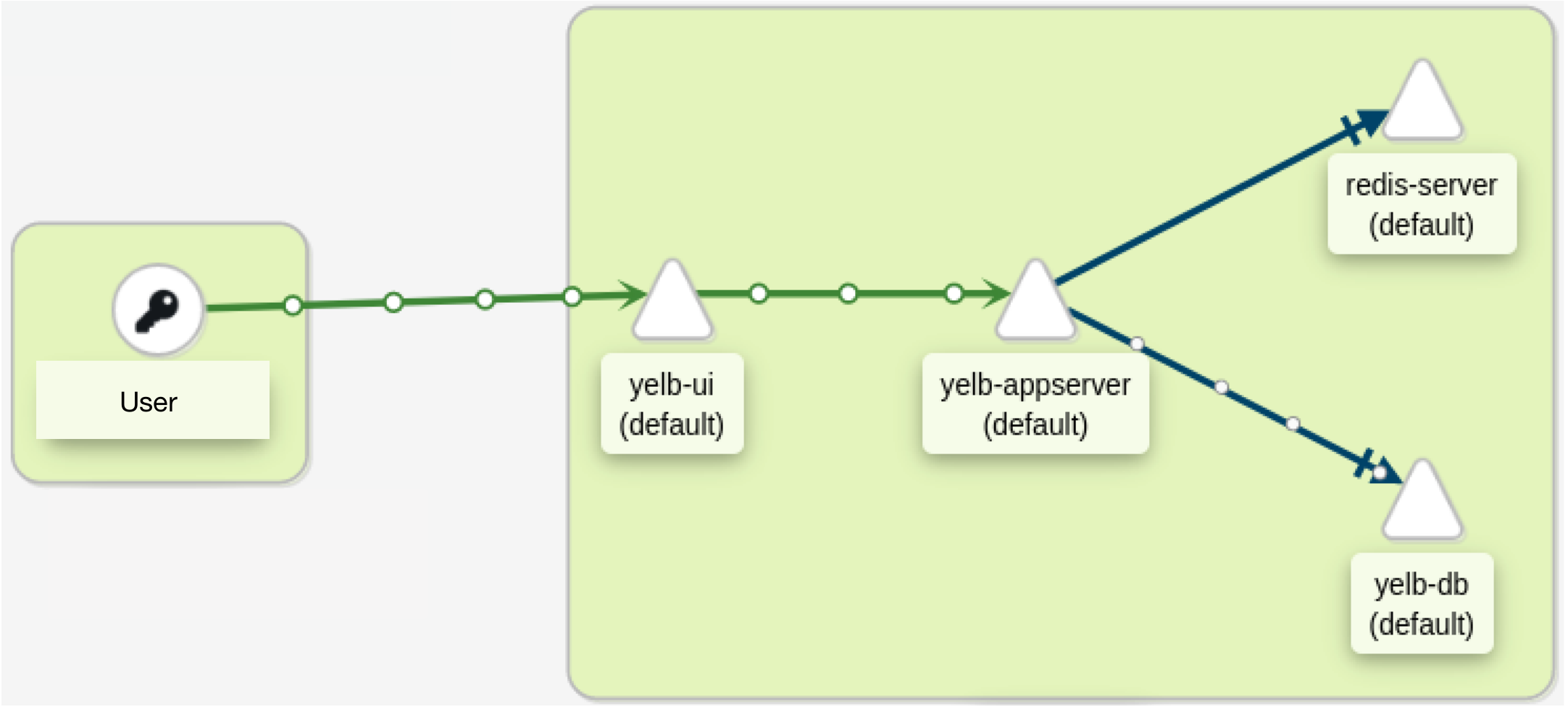}
	\vspace*{-2ex}
    \caption{\label{fig:yelb}Istio view of the service graph for the Yelb case study}
\end{figure}

 \head{Yelb Application}
The Yelb application consists of four services: a user interface service, an application server (appserver) service, a redis server service, and a database service. 
It allows users to vote for their preferred restaurant among given options, and updates a pie chart based on the number of votes received for each option. 
The Istio view of the service graph for the Yelb application is shown in Figure~\ref{fig:yelb}.

\subsection{Test Scenarios for Fault Injection} 

\noindent
Chaos Engineering allows the injection of two types of faults in a running application: infrastructure level and functional level.
In the context of microservices-based applications deployed in a CDC cluster, the infrastructure level faults include the faults that occur due to issues with CDC configurations and resources.
These faults can be mostly injected using predefined functions available within various chaos toolkits.
The functional level faults are unique to the application and require some additional knowledge of the application's backend logic and data flow.
Chaos Engineering provides support for functional level faults, which can be injected by writing custom probes and action methods for a specific scenario and calling them in a chaos experiment script. 

The artifact presents 5 fault injection scenarios, 4 using the smart office case study and 1 using the Yelb app.

\begin{enumerate}[label=\textit{{FS-\arabic*:}}, ref=FS-\arabic*, left=\parindent]\itshape
	\item \textit{a deployed sensor is down unexpectedly.}
	\item \textit{a deployed sensor sends erroneous readings.}
	\item \textit{a running service is down abruptly.}
	\item \textit{a running service is delayed.}
    \item \textit{a running service is loaded with a high service request rate (SRR).}
\end{enumerate}
Table~\ref{tab:FI} presents an overview of faults injected with respect to the fault level and the target application. 
The infrastructure level fault injection is executed on the system service in a black-box manner.
The infrastructure \& functional level fault execution is performed in a grey-box manner where some additional access to the service's functional status is required.

\begin{table}
	\vspace*{1ex}
    \caption{\label{tab:FI} Test Scenarios for Fault Injection}
	\centering
	\begin{tabularx}{0.95\columnwidth}{lll}
		\toprule
        Fault Injection &    Fault Level  &  Target Application   \\
        \midrule
		Sensor Down &   Infrastructure \& Functional   & Smart-Office \\
		Sensor Faulty & Infrastructure \& Functional & Smart-Office \\
		Service Delayed & Infrastructure & Smart-Office \\
        Service Down & Infrastructure & Smart-Office \\
		High SRR &  Infrastructure & Yelb-App  \\
		\bottomrule
	\end{tabularx} 
    \vspace*{-2ex}
\end{table}

\section{Experiments}
\label{sec:use_artifact}

\begin{figure}[b]
	\vspace*{-1ex}
	\centering
	\includegraphics[width=0.5\columnwidth]{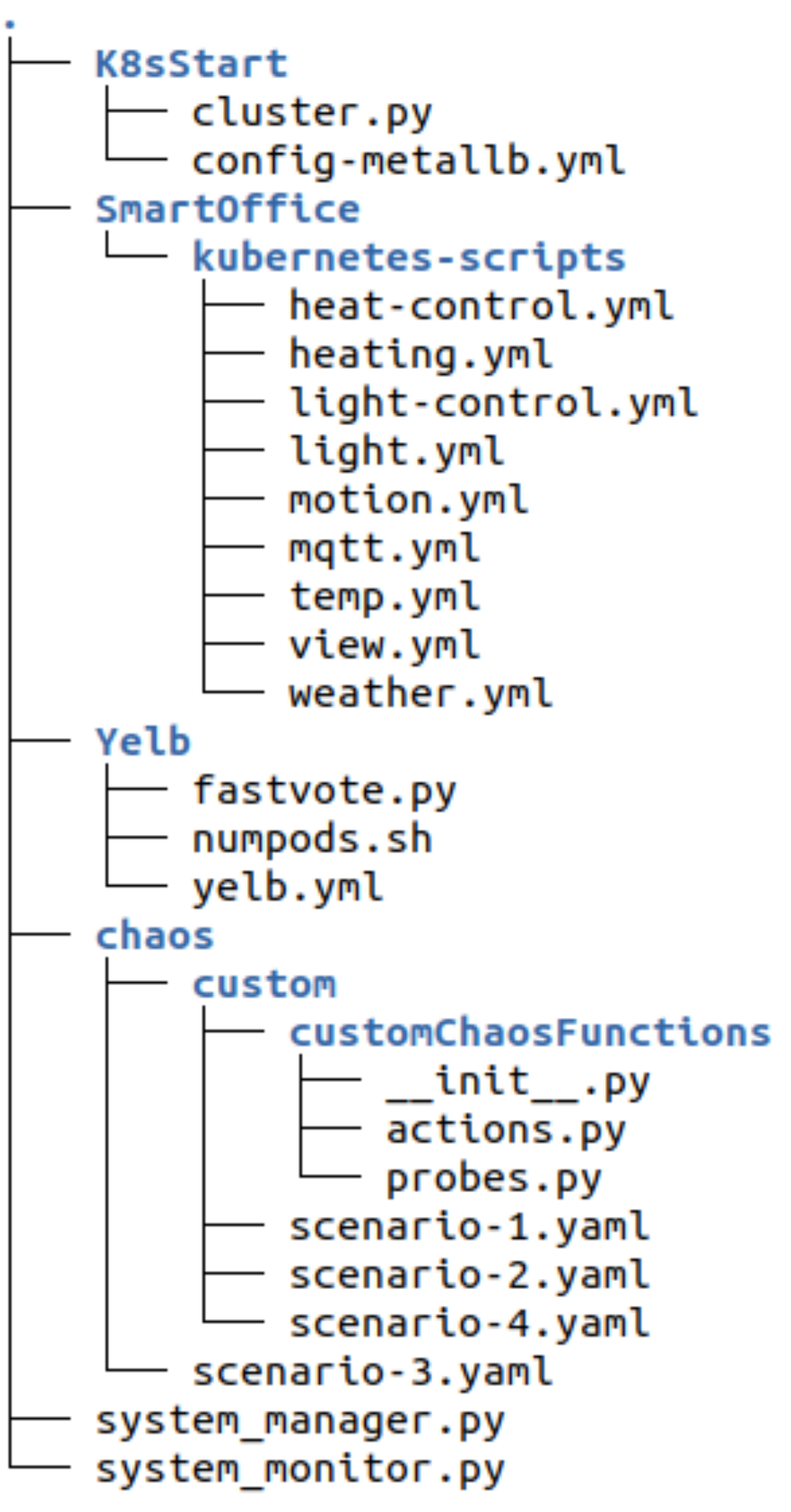}
	\caption{\label{fig:dir}Artifact's directories' hierarchy}
\end{figure}

\noindent
The artifact is available in a virtual machine.
The directory structure for artifact implementation is shown in Figure~\ref{fig:dir}.

\subsection{How to use the artifact: Smart-Office Scenarios}

\noindent
There are four main steps, along with two optional steps for online data visualization, to run the artifact for the smart-office scenarios (refer to Table~\ref{tab:runSO}).

The first step is to create or start the K8s cluster by running the script \textit{cluster.py}. 
This will create a new minikube-based K8s cluster with 4 CPUs and 2000 MB of memory. 
The newly created cluster will be configured with metalLB load balancer and have istio mesh installed, with istio-injection enabled and monitoring tools installed.
Next, you can deploy the demo application's services for a specified scenario by running the script \textit{system\_monitor.py X Y}. 
The value for X represents the scenario number and can be $1$, $2$, $3$, or $4$ to run FS-1, FS-2, FS-3, or FS-4, respectively. 
The value for Y is either $0$ if you want to run the system monitor alone, or $1$ if you want to run the system monitor with the system manager to recover the services from failures.
To view the running deployments and services in the k8s cluster, run the command: \textit{kubectl get all}. 
Step three involves activating the virtual chaos environment for chaos injection. 
Then, run the chaos command \textit{chaos run scenario-X.yml} to inject a failure corresponding to the running scenario. 
Where, \textit{scenario-X} represents the chaos experiment to be injected, and X can be $1$, $2$, $3$, or $4$ for FS-1, FS-2, FS-3, or FS-4, respectively.
Finally, you can use online monitoring tools to examine the online monitoring data for service statuses and traffic flow. 
If desired, the Kiali dashboard and Grafana dashboard can be launched in separate terminals to observe the service data metrics.

{\renewcommand{\arraystretch}{1.5}%
\begin{table}
	\caption{\label{tab:runSO} Run Smart-Office Scenarios}
	\centering
	\begin{tabularx}{0.79\columnwidth}{l}
        \midrule
        Step 1: Start K8s Cluster \\
		 \ \ \ \ \ \ \ \ \ \ \ \ \ \ \verb|python3 cluster.py| \\
		Step 2: Run Demo-App Scenario \\
         \ \ \ \ \ \ \ \ \ \ \ \ \ \ \verb|python3 system_monitor.py| \textit{X} \textit{Y} \\
        Step 3: Run Chaos Virtual Env \\
         \ \ \ \ \ \ \ \ \ \ \ \ \ \ \verb|~/.venvs/chaosEnv/bin/activate|  \\
		Step 4: Run Chaos Script \\
         \ \ \ \ \ \ \ \ \ \ \ \ \ \ \verb|chaos run scenario-X.yaml| \\
        Step 5: Kiali visualization [Optional]\\
         \ \ \ \ \ \ \ \ \ \ \ \ \ \ \verb|istioctl dashboard kiali| \\
        Step 6: Grafana visualization [Optional]\\
         \ \ \ \ \ \ \ \ \ \ \ \ \ \ \verb|istioctl dashboard grafana| \\
		\bottomrule
	\end{tabularx} 
    \vspace*{-2ex}
\end{table}
}

\subsection{How to use the artifact: Yelb-App Scenario}

\noindent
The execution process for the Yelb application scenario involves four main steps, with two additional optional steps for online data visualization, as described in Table~\ref{tab:runYelb}.

The first step is to create the K8s cluster, followed by the deployment of demo application services. 
The app-server service can be enabled for auto-scaling using the horizontal pod auto-scaling (HPA) functionality of K8s by running the following command: \textit{kubectl autoscale deployment yelb-appserver --cpu-percent=10 --min=1 --max=20}. 
The parameters \textit{cpu-percent}, \textit{min}, and \textit{max} represent the dedicated CPU percentage, the minimum number of replicas to be running, and the maximum number of replicas that can run under heavy load, respectively. These values can be adjusted based on the available resources in the cluster.
In the third step, the pods' logger is executed by running the shell script \textit{numpods.sh}. 
This script records changes in the number of replicas over time and writes the results to a log file, which can later be analyzed to understand the relationship between the number of active users and the usage of CPU resources, and the increase or decrease in replicas.
The fourth step is to run the scripted chaos for the fast voting onto the yelb-appserver. 
Finally, for real-time data observation, the Kiali dashboard and Grafana dashboard can be launched in separate terminals, as an optional step.

{\renewcommand{\arraystretch}{1.5}%
\begin{table}
	\caption{\label{tab:runYelb} Run Yelb-App Scenario}
	\centering
	\begin{tabularx}{0.75\columnwidth}{l}
        \midrule
        Step 1: Start K8s Cluster \\
		 \ \ \ \ \ \ \ \ \ \ \ \ \ \ \verb|python3 cluster.py| \\
		Step 2: Run Demo-App Scenario \\
         \ \ \ \ \ \ \ \ \ \ \ \ \ \ \verb|python3 system_monitor.py| \textit{X} \verb|0| \\
        Step 3: Run pods logger \\
         \ \ \ \ \ \ \ \ \ \ \ \ \ \ \verb|./numpods.sh| \\
		Step 4: Start overloading yelb-app \\
         \ \ \ \ \ \ \ \ \ \ \ \ \ \ \verb|python3 fastvote.py| <url>\\
        Step 5: Kiali visualization [Optional]\\
         \ \ \ \ \ \ \ \ \ \ \ \ \ \ \verb|istioctl dashboard kiali| \\
        Step 6: Grafana visualization [Optional]\\
         \ \ \ \ \ \ \ \ \ \ \ \ \ \ \verb|istioctl dashboard grafana| \\
		\bottomrule
	\end{tabularx} 
    \vspace*{-2ex}
\end{table}
}

\subsection{Results}

\begin{figure}[t]
	\vspace*{1ex}
	\centering
	\includegraphics[width=0.98\columnwidth]{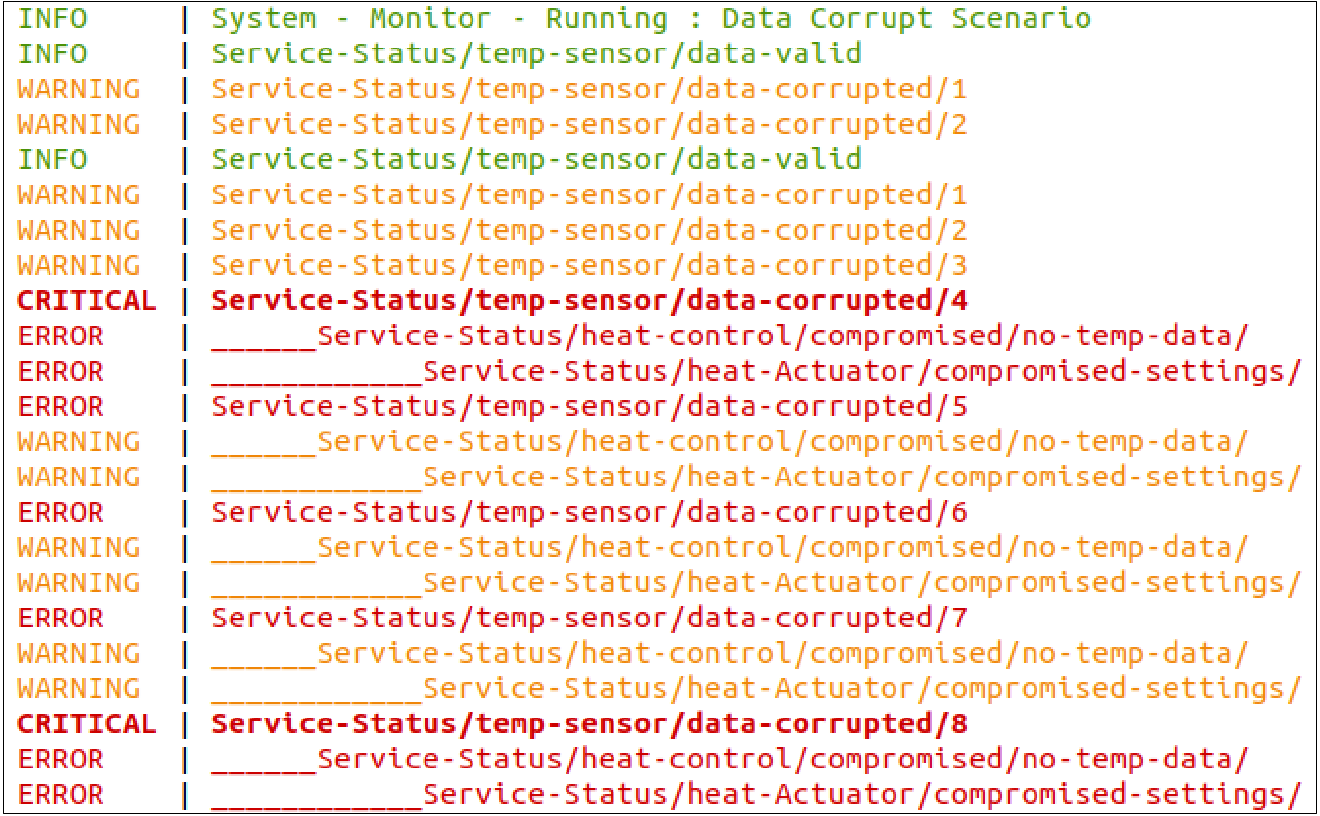}
	\vspace*{-1ex}
	\caption{\label{fig:Sce10}Running scenario 1 without system manager service}
\end{figure}

\begin{figure}[t]
	\vspace*{1ex}
	\centering
	\includegraphics[width=0.75\columnwidth]{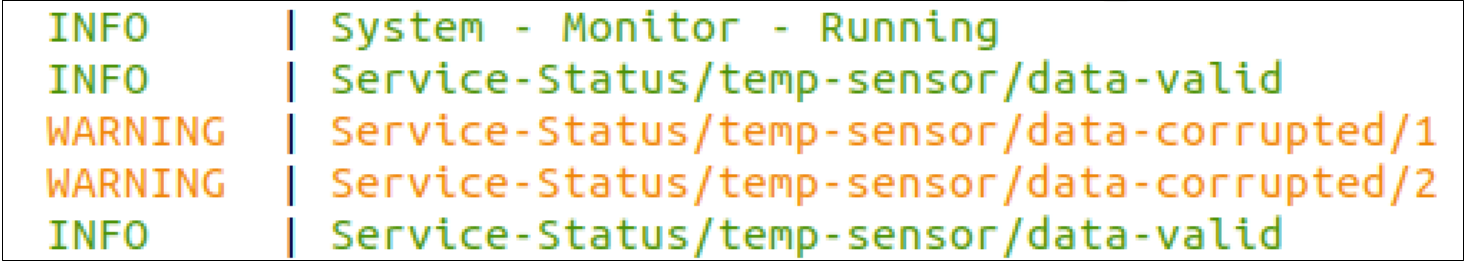}
	\vspace*{-1ex}
	\caption{\label{fig:Sce11}Running scenario 1 with system manager service}
	\vspace*{-1ex}
\end{figure}

\noindent
\head{Smart-Office Scenarios}
The first four scenarios (FS1, FS2, FS3, and FS4) are designed to evaluate the effect of chaos injection and failures on a running system, both with and without the deployment of a system management service. 
They provide a comparison of the system's behavior when it fails to recover from an injected fault and when it successfully recovers to evaluate the system's behavior and recovery capabilities.
 Figure~\ref{fig:Sce10} depicts the system monitor output for FS-1 when run without a management service (i.e., running \textit{python3 system\_monitor 1 0}). 
 In this scenario, an injected fault in the form of data corruption leads to a cascading failure that the system is unable to recover. 
 On the other hand, Figure~\ref{fig:Sce11} shows the system monitor output for FS-1 when run with a management service (i.e., running \textit{python3 system\_monitor 1 1}). 
 In this case, the system manager is able to quickly identify and recover from the failed service, resulting in a much smoother and more efficient recovery process.

\begin{figure}
	\centering
	\includegraphics[width=0.9\columnwidth]{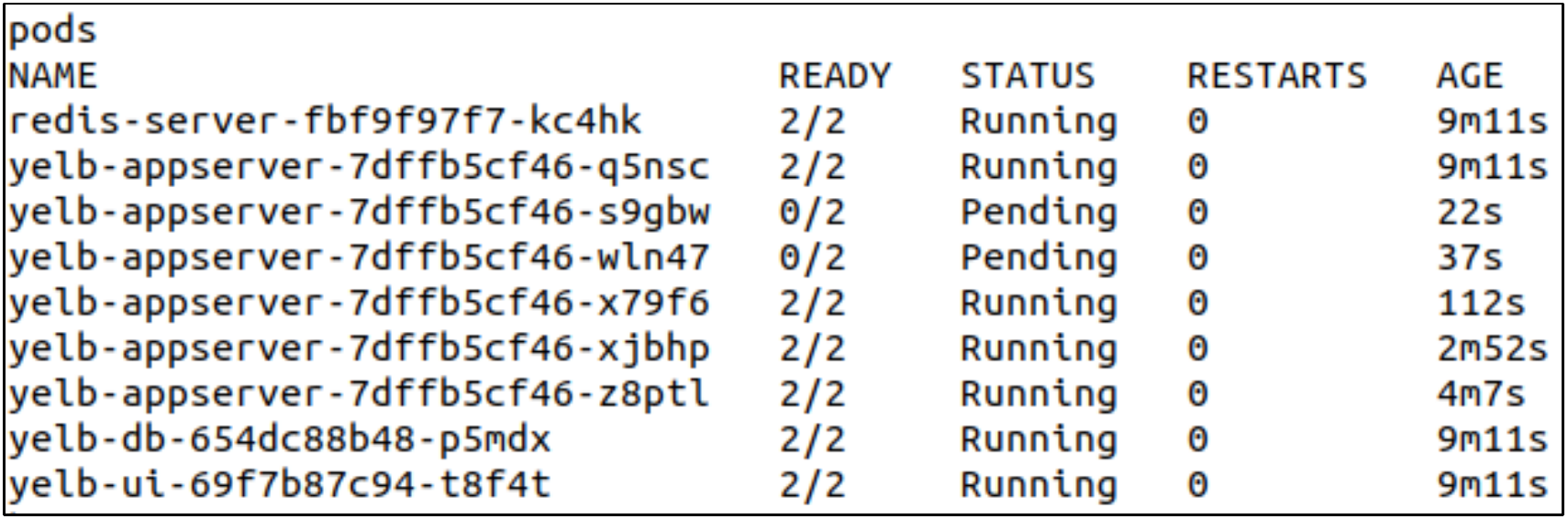}
	\vspace*{-1ex}
	\caption{\label{fig:Sce5-1} Pods data view in logfile}
\end{figure}

\begin{figure}
	\centering
	\includegraphics[width=\columnwidth]{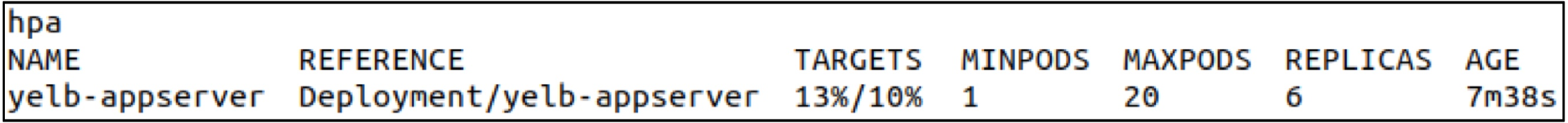}
	\vspace*{-3ex}
	\caption{\label{fig:Sce5-2} HPA data view in logfile}
\end{figure}

\begin{figure}
	\vspace*{1ex}
	\centering
	\includegraphics[width=0.55\columnwidth]{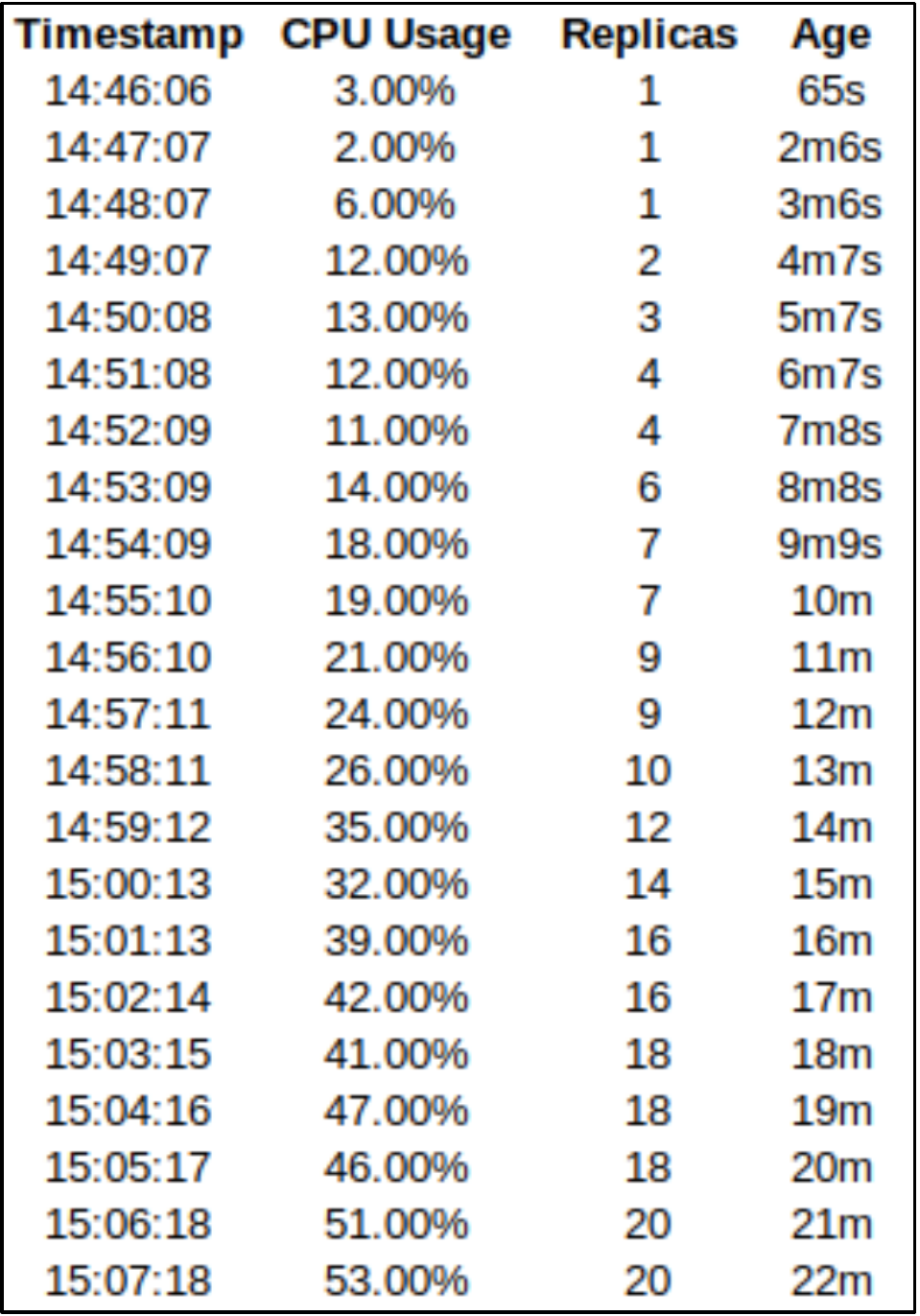}
	\vspace*{-1ex}
	\caption{\label{fig:Sce5-3} Data extracted from numpods logfile}
	\vspace*{-1ex}
\end{figure}

\head{Yelb-App Scenario}
The FS5 evaluates the scalability of services deployed in a Kubernetes cluster by imposing different loads on the service.
The resulting log file captures two types of data:  information about all running pods (obtained through \textit{kubectl get pods}) and data on HPA deployment (obtained through \textit{kubectl get hpa}).
The log is updated every 30 seconds and records the name, status, number of restarts, and age of each running pod (as shown in Figure~\ref{fig:Sce5-1}), and the name, CPU targets, minimum and maximum number of pods, number of replicas, and age of the HPA deployment (as shown in Figure~\ref{fig:Sce5-2}).
This log can be analyzed to uncover patterns in the service's performance and resource utilization over time. 
For example, Figure~\ref{fig:Sce5-3} presents sample data extracted from observing generated log files.
The extracted data depicts a dramatic increase in CPU usage and the number of replicas over a 22-minute period.

\section{Concluding Remarks}
\label{sec:conclusion}

\head{Contributions}
This paper presents an artifact that provides a detailed overview of 
how the \emph{CHESS} approach can be used to evaluate a system's 
resilience and ability to recover from various types of faults. 
The implemented modules highlight the effectiveness of the system monitoring and system managing services in detecting and mitigating failures in the system. 
The artifact consists of (i) predefined
functional and infrastructural level fault injection scenarios,
(ii) a self-monitoring service that presents extensive logs
for the deployed services’ normal and abnormal behaviors,
(iii) the managing system service that reacts to the system
abnormal behavior traces and brings the system back to the
stable condition, and (iv) a comparison of the service failure
and cascading effects with and without deployment of the
managing system service.
The artifact is available on Zenodo,$^{\ref{artifact}}$ and a demo video is on YouTube.\footnote{~Artifact demo video: \url{https://youtu.be/CBcaPJgpi-o}}

\head{Applicability}
The artifact provides the SEAMS community with support for one of the critical tasks of software engineering research, i.e., the \emph{systematic evaluation} of novel approaches.
It aligns with the growing desire to produce self-adaptation artifacts that support industry-relevant research~\cite{weyns2022:guidelines}
by using chaos engineering to systematically observe containerized applications in Docker~\cite{simonsson2021:observability}.
Artifact components are reusable, extendable, and modifiable for new case studies. 
Existing fault scenarios can be combined and expanded to create complex ones. 
The custom fault injection scripts can inspire exploration of grey-box level fault injection and testing.

\head{Future work}
Directions of interest include exploring the use of observability data and system logs for chaos engineering-controlled data synthesis. 
In addition, techniques for the automated selection of regions for chaos experiments 
based on the health and performance status of services are also of interest. 
Finally, the inclusion of contextual information, such as ontologies or knowledge graphs, 
can also be considered to enhance fault injection targeting.

\medskip

\head{Acknowledgments}
This research is supported by the Research Council of Norway 
through the 
cureIT project (\#300461)
and used resources provided by 
the Experimental Infrastructure for Exploration of Exascale Computing (eX3), 
supported by the Research Council of Norway (\#270053)%
.

\newpage  %
\enlargethispage{-1.7cm} %
\printbibliography 

\end{document}